\documentstyle[12pt,epsfig,axodraw,a4]{article} 
\newcommand{\vs}{\vspace{-0.25cm}}
\begin{document} 

\begin{center}
{\Large{\bf $\Sigma$-nuclear spin-orbit coupling from two-pion exchange}}  
\bigskip

N. Kaiser\\
\medskip
{\small Physik-Department T39, Technische Universit\"{a}t M\"{u}nchen,
    D-85747 Garching, Germany}
\end{center}
\medskip
\begin{abstract}
Using SU(3) chiral perturbation theory we calculate the density-dependent
complex-valued spin-orbit coupling strength $U_{\Sigma ls}(k_f)+ i\, W_{\Sigma 
ls}(k_f)$ of a $\Sigma$ hyperon in the nuclear medium. The leading long-range  
$\Sigma N$ interaction arises from iterated one-pion exchange with a 
$\Lambda$ or a $\Sigma$ hyperon in the intermediate state. We find from this 
unique long-range dynamics a sizeable ``wrong-sign'' spin-orbit coupling 
strength of $U_{\Sigma ls}(k_{f0}) \simeq -20$\,MeVfm$^2$ at normal nuclear
matter density $\rho_0 = 0.16\,$fm$^{-3}$. The strong $\Sigma N\to \Lambda N$ 
conversion process contributes at the same time an imaginary part of $W_{
\Sigma ls}(k_{f0}) \simeq -12$\,MeVfm$^2$. When combined with estimates of the 
short-range contribution the total $\Sigma$-nuclear spin-orbit coupling becomes
rather weak. 
\end{abstract}

\bigskip


Hypernuclear physics has a long and well-documented history \cite{dover,
chrien,dovgal}. One primary goal in this field is to determine from the 
experimental data the nuclear mean-field potentials relevant for the 
hyperon single-particle motion. For the $\Lambda$ hyperon the situation is 
by now rather clear and the following quantitative features have emerged. The 
attractive nuclear mean-field potential for a $\Lambda$ hyperon is about half 
as strong as the one for nucleons in nuclei: $U_\Lambda \simeq -28\,$MeV 
\cite{2}. With this value of the potential depth the empirical single-particle 
energies of a $\Lambda$ bound in hypernuclei are well described over a wide 
range in mass number. On the other hand, the $\Lambda$-nucleus spin-orbit 
interaction is found to be extraordinarily weak. For example, recent precision 
measurements \cite{ajimura} of $E1$-transitions from $p$- to $s$-shell 
orbitals in $^{13}_\Lambda C$ give a $p_{3/2}-p_{1/2}$ spin-orbit splitting of 
only $(152\pm 65)\,$keV to be compared with a value of about $6$\,MeV in 
ordinary $p$-shell nuclei. 

In case of the $\Sigma$ hyperon recent developments 
have lead to a revision concerning the sign and magnitude of its nuclear 
mean-field potential \cite{galneu}. Whereas an earlier analysis of the shifts 
and widths  of x-ray transitions in $\Sigma^-$ atoms came up with an 
attractive (real) $\Sigma$-nucleus optical potential of about $-27\,$MeV 
\cite{dover}, there is currently good experimental and phenomenological 
evidence for a substantial $\Sigma$-nucleus repulsion. A reanalysis of the 
$\Sigma^-$ atom data in Ref.\cite{batty} including the then available precise 
measurements of W and Pb atoms and employing phenomenological
density-dependent fits has lead to a $\Sigma$-nucleus potential with a 
strongly repulsive core (of height $\sim 95\,$MeV) and a shallow attractive 
tail outside the nucleus. The inclusive $(\pi^-,K^+)$ spectra on 
medium-to-heavy nuclear targets measured at KEK \cite{noumi,saha} give more 
direct evidence for a strongly repulsive $\Sigma$-nucleus potential. In the 
framework of the distorted wave impulse approximation, a best fit of the  
measured $(\pi^-,K^+)$ inclusive spectra on Si, Ni, In and Bi targets is 
obtained with a $\Sigma$-nucleus repulsion of about $90\,$MeV. However, the 
detailed description of the $\Sigma^-$ production mechanism plays an important 
role for the extracted value of the $\Sigma$-nucleus repulsion. Within a 
semiclassical  distorted wave model \cite{kohno}, which avoids the  
factorization approximation by an averaged differential cross section, the KEK 
data can also be well reproduced with a complex $\Sigma$-nucleus potential of 
strength $(30- 20\,i)$\,MeV. Concerning the $\Sigma$-nucleus spin-orbit 
coupling there exist so far no experimental hints for it. Most theoretical 
models \cite{pirner,bouyssy} predict the $\Sigma$-nucleus spin-orbit coupling
to be strong (i.e. comparable to the one of nucleons). The basic argument for 
a strong spin-orbit coupling is provided by the large and positive value of 
the tensor-to-vector coupling ratio of the $\omega$ meson to 
the $\Sigma$ hyperon assuming vector meson dominance and the non-relativistic 
quark model with SU(6) spin-flavor symmetry. The G-matrix calculations by the 
Kyoto-Niigata group \cite{fuji} using the hyperon-nucleon interaction as 
derived from their SU(6) quark model predict a $\Sigma$-nucleus spin-orbit 
coupling which is about half as strong as the one of nucleons. However, due to 
the presence of the strong $\Sigma N\to \Lambda N$ conversion process in the 
nuclear medium one expects the $\Sigma$-nucleus spin-orbit coupling strength  
to have also an imaginary part. This possibility has generally been ignored in 
quark and one-boson exchange models. 

Recently, we have applied chiral effective field theory to calculate the 
hyperon mean-fields in nuclear matter \cite{lambdapot}. In this approach the 
small $\Lambda$-nuclear spin-orbit interaction finds a novel explanation 
in terms of an almost complete cancellation between short-range contributions 
(estimated from the known nucleonic spin-orbit coupling strength) and 
long-range terms generated by iterated one-pion exchange with intermediate
$\Sigma$ hyperons. The exceptionally small $\Sigma\Lambda$ mass splitting of
$M_\Sigma -M_\Lambda =77.5\,$MeV influences hereby prominently the effect 
coming from the second order $1\pi$-exchange tensor interaction. Furthermore, 
it has been shown in Ref.\cite{jorge} that the proposed cancellation mechanism
does not get disturbed by the inclusion of analogous two-pion exchange 
processes involving decuplet baryons ($\Delta(1232)$ and $\Sigma^*(1385)$) in
the intermediate state with considerably larger mass splittings. The 
density-dependent complex $\Sigma$-nuclear mean-field  $U_\Sigma(k_f)+ i\, 
W_\Sigma(k_f)$ has also been calculated in the same framework in 
Ref.\cite{sigmapot}. It has been found that genuine long-range\footnote{
Genuine long-range means that (unique) part of the pion-loop which depends 
exclusively on small scales ($k_f, m_\pi, \Delta$), but not any high-momentum 
cutoff. In case of the $\Sigma$-nuclear mean field $U_\Sigma(k_f)$ it seems 
that the net short-range contribution is small \cite{sigmapot}. For the 
$\Lambda$ single-particle potential $U_\Lambda(k_f)$ an attractive short-range 
contribution \cite{lambdapot} is however necessary in order to reproduce the
empirical potential depth of $-28\,$MeV. A deeper understanding of this
feature is presently missing.}
contributions from iterated one-pion exchange with intermediate $\Lambda$ and
$\Sigma$ hyperons sum up to a moderately repulsive (real) single-particle 
potential of $U_\Sigma(k_{f0}) \simeq 59\,$MeV at normal nuclear matter 
density $\rho_0 = 0.16\,$fm$^{-3}$. The $\Sigma N\to \Lambda N$ conversion 
process induced by one-pion exchange generates at the same time an imaginary 
single-particle potential of $W_\Sigma(k_{f0}) \simeq -21.5$\,MeV. This value
is in fair agreement with empirical determinations \cite{batty} and quark model
predictions \cite{kohno2}. 
The purpose of the present short paper is to calculate in the same chiral 
effective field theory framework the density-dependent complex-valued 
$\Sigma$-nuclear spin-orbit coupling strength. As for the $\Lambda$ hyperon 
\cite{lambdapot} we do find a sizeable ``wrong-sign'' spin-orbit coupling from 
the second-order one-pion exchange tensor interaction. When combined with 
estimates of the short-range contribution (employing QCD sum rule predictions) 
the total $\Sigma$-nuclear spin-orbit coupling becomes rather weak.   

Let us begin with some basic considerations. The pertinent quantity to extract
the $\Sigma$-nuclear spin-orbit coupling is the spin-dependent part of the
self-energy of a $\Sigma$ hyperon interacting with weakly inhomogeneous
isospin-symmetric (spin-saturated) nuclear matter. Let the $\Sigma$ hyperon
scatter from initial momentum $\vec p- \vec q/2$ to final momentum $\vec p+ 
\vec q/2$. The spin-orbit part of the self-energy is then:  
\begin{equation}\Sigma_{\rm spin} = {i \over 2} \,\vec \sigma \cdot (\vec q
\times \vec p\,) \, \Big[ U_{\Sigma ls}(k_f)+i\, W_{\Sigma ls}(k_f)\Big] \,,
\end{equation}
where the density-dependent spin-orbit coupling strength $U_{\Sigma ls}(k_f)
+i\,W_{\Sigma ls}(k_f)$ is taken in the limit of homogeneous nuclear matter 
(characterized by its Fermi momentum $k_f$) and zero external 
$\Sigma$-momenta: $\vec p =\vec q =0$. The more familiar spin-orbit 
Hamiltonian follows from Eq.(1) by multiplication with a density form factor
and Fourier transformation $\int d^3 q \exp(i \vec q \cdot \vec r\,)$. For 
orientation, consider first the $\omega$ meson exchange between the $\Sigma$ 
hyperon and the nucleons. The non-relativistic expansion of the vector 
(and tensor) coupling vertex between Dirac spinors of the $\Sigma$ hyperon 
gives rise to a spin-orbit term proportional to $i\,\vec \sigma \cdot (\vec q 
\times \vec p\,)/4M_\Sigma^2$. Next one takes the limit of homogeneous nuclear 
matter (i.e. $\vec q=0$), performs the remaining integral over the nuclear 
Fermi sphere and arrives at the familiar result:  
\begin{equation}  U_{\Sigma ls}(k_f)^{(\omega)} = {g_{\omega \Sigma}(1+2
\kappa_{\omega \Sigma}) g_{\omega N} \over 2M_\Sigma^2 m_\omega^2} \, \rho \,,
\end{equation} 
linear in density $\rho= 2k_f^3/3\pi^2$. Here, $\kappa_{\omega\Sigma}$ denotes
the tensor-to-vector coupling ratio of the $\omega$ meson to the $\Sigma$ 
hyperon.
\begin{figure}
\begin{center}
\includegraphics[scale=1.]{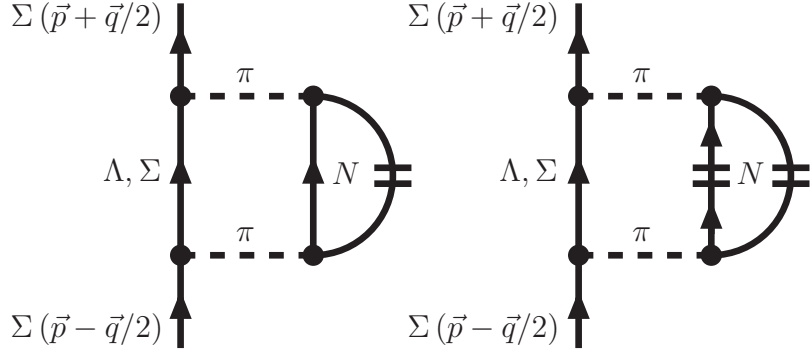}
\end{center}
\vspace{-.4cm}
\caption{Iterated one-pion exchange diagrams with $\Lambda$ and $\Sigma$ 
hyperons in the intermediate state generating a $\Sigma$-nuclear spin-orbit
coupling. The horizontal double-line symbolizes the filled Fermi sea of
nucleons, i.e. the medium insertion $-\theta(k_f-|\vec p_j|)$ in the in-medium
nucleon propagator.}
\vspace{-.3cm}
\end{figure}

The crucial observation is now that the (left) iterated one-pion exchange 
diagram in Fig.\,1 generates also a (sizeable) spin-orbit coupling term. 
The prefactor ${i\over2}\vec\sigma \times \vec q$ is immediately identified by 
rewriting the product of $\pi\Sigma B$-interaction vertices $\vec\sigma\cdot(
\vec l-\vec q/2)\,\vec \sigma \cdot (\vec l + \vec q/2) = {i\over 2}(\vec 
\sigma \times \vec q\,)\cdot (-2 \vec l\,) +\dots $ at the open baryon line. 
For all remaining parts of the diagram one can then take the limit of 
homogeneous nuclear matter (i.e. $\vec q=0$). The other essential factor 
$\vec p$ comes from the energy denominator $-\Delta^2+\vec l\cdot(\vec l-\vec 
p_1+\vec p\,)$. The $\Sigma\Lambda$ mass splitting is rewritten here in terms 
of the small scale parameter $\Delta = \sqrt{M_B (M_\Sigma -M_\Lambda)} \simeq 
285\,$MeV with $M_B=(2M_N+M_\Lambda+M_\Sigma)/4\simeq 1047\,$MeV a mean baryon 
mass. It serves the purpose to average out small differences in the kinetic
energies of the various baryons involved. Keeping only the term linear in the 
external momentum $\vec p$ one finds from the left diagram in Fig.\,1 with a 
$\Lambda$ hyperon in the intermediate state the following contribution to the
$\Sigma$-nuclear spin-orbit coupling strength:
\begin{eqnarray} U_{\Sigma ls}(k_f)^{(2\pi\Lambda)} + i\, W_{\Sigma ls}(k_f)^{
(2\pi\Lambda)}  &=& - {2D^2 g_A^2 \over 9f_\pi^4}\! \int\limits_{|\vec p_1| < 
k_f} \!\!\!{d^3 p_1 d^3 l\over (2\pi)^6} { M_B\, {\vec l}\,^4 \over (m_\pi^2 
+{\vec l}\,^2 )^2\, [ -\Delta^2 -i0+{\vec l}\,^2 -\vec l \cdot \vec p_1]^2 } 
\nonumber \\ &=&  {2\over 3} {\partial \over \partial \Delta^2} \Big[U_\Sigma(
k_f)^{(2\pi\Lambda)}+i\, W_\Sigma(k_f)^{(2\pi\Lambda)} \Big] \,. \end{eqnarray}
Here, $D=0.84$ and $F=0.46$ \cite{lambdapot} denote the SU(3) axial vector
coupling constants together with $g_A = D+F =1.3$ the nucleon axial vector
coupling constant. $f_\pi = 92.4\,$MeV is the pion decay constant and 
$m_\pi=138\,$MeV the average pion mass. Note that the loop integral in Eq.(3) 
is convergent as its stands. Most useful is actually the
representation of the spin-orbit coupling strength as a derivative of the
$\Sigma$-nuclear potential $U_\Sigma(k_f)+i\, W_\Sigma(k_f)$ with respect to 
the (mass splitting) parameter $\Delta^2$. Using the analytical expressions in
Ref.\cite{sigmapot} to evaluate this derivative we find for the real and 
imaginary part:  
\begin{equation}  U_{\Sigma ls}(k_f)^{(2\pi\Lambda)}  = {D^2 g_A^2 M_B m_\pi^2
\over 72 \pi^3 f_\pi^4} \bigg\{ (4+2\delta) \arctan{ \sqrt{u}\over 1+\delta}
- {3u+( 1+\delta)(4+2\delta)\over u +(1+\delta)^2} \sqrt{u} \bigg\} \,, 
\end{equation} 
\begin{eqnarray}  W_{\Sigma ls}(k_f)^{(2\pi\Lambda)} &=& {D^2 g_A^2 M_B m_\pi^2
\over 72 \pi^3 f_\pi^4} \Bigg\{- {u+(1+\delta)(2+\delta)\over u +(1+\delta)^2} 
\sqrt{u(4\delta+u)}\nonumber \\ && +(4+2\delta) \ln{u+2+2\delta+\sqrt{u(4
\delta+u)} \over 2[ u +(1+\delta)^2]^{1/2}}  \Bigg\} \,,  \end{eqnarray}
with the abbreviations $u= k_f^2/m_\pi^2$ and $\delta = \Delta^2/m_\pi^2$. The
right diagram in Fig.\,1 with two medium insertions represents the Pauli 
blocking correction. In comparison to the expression in Eq.(3) the sign is
reverse and the momentum transfer $\vec l$ gets replaced by $\vec l = \vec p_1
- \vec p_2$ with $\vec p_2$ to be integrated over a  Fermi sphere of radius 
$k_f$, i.e. $|\vec p_2| < k_f$. In case of the real part one is left with a 
double-integral of the form: 
\begin{eqnarray}U_{\Sigma ls}(k_f)^{(2\pi\Lambda)}_{\rm Pauli} &=&{D^2g_A^2M_B 
m_\pi^2\over 36\pi^4  f_\pi^4} -\!\!\!\!\!\!\int_0^u \!\!dx \int_0^u \!\!dy \,
{1 \over (2\delta+1+x-y)^2 } \,  \Bigg\{ {(2\delta+x-y)^2  \sqrt{xy}\over 2(
\delta-y)^2-2x y} \nonumber\\ &&+ {2 \sqrt{xy} \over (1+x+y)^2-4x y} +{2\delta 
+x-y \over 2\delta+1+x-y} \ln{|\delta - y -\sqrt{xy} | (1+x+y - 2\sqrt{xy})
\over |\delta - y +\sqrt{xy}| (1+x+y + 2\sqrt{xy})}\Bigg\} \,, \nonumber \\
\end{eqnarray}
where the first term in brackets has to be treated as a principal value
integral. In practice this is done by solving the $\int_0^u\! dx$-integral 
analytically and converting the occurring logarithms into logarithms of
absolute values. The Pauli blocking correction to the imaginary part 
$W_{\Sigma ls}(k_f)$ can even be written in closed analytical form:  
\begin{eqnarray}W_{\Sigma ls}(k_f)^{(2\pi\Lambda)}_{\rm Pauli}& = &{D^2g_A^2
M_B m_\pi^2\over 72\pi^3  f_\pi^4}\, \theta(\sqrt{2}k_f-\Delta)  \Bigg\{
{u\over 2}- \delta -1 +{1 \over 1+2 \delta}+ { u \delta \over u+\delta^2}  
\nonumber\\ &&+ {u (1-\delta) \over 2u+2(1+\delta)^2}+ {u+(1+\delta)(2+\delta)
\over 2u +2(1 +\delta)^2} \sqrt{u(4 \delta+u)} +2\ln(2+4 \delta) \nonumber\\
&& + \delta \ln(2+2 \delta^2 u^{-1}) -(2+\delta)
\ln\Big[u+2+2\delta+\sqrt{u(4\delta+u)}\,\Big]  \Bigg\} \,.  \end{eqnarray}
Interestingly, there is a threshold condition $k_f>\Delta/\sqrt{2}$ for Pauli 
blocking to become active in the imaginary part. The threshold opens at about 
one half of nuclear matter saturation density $\rho_{\rm th}= 0.072\,{\rm fm
}^{-3} =0.45\rho_0$.  

The additional contributions from the iterated one-pion exchange diagrams with 
a $\Sigma$ hyperon in the intermediate state are obtained by substituting 
axial vector coupling constants, $D^2\to 6 F^2$, and dropping the $\Sigma
\Lambda$ mass splitting, $\delta \to 0$. The explicit expressions for these
contributions to the complex $\Sigma$-nuclear spin-orbit coupling strength 
read:
\begin{equation}  U_{\Sigma ls}(k_f)^{(2\pi\Sigma)}  = {F^2 g_A^2 M_B m_\pi^2
\over 12 \pi^3 f_\pi^4} \bigg\{ 4 \arctan \sqrt{u} - {4+3u \over 1+u} \sqrt{u}
\bigg\} \,, \end{equation} 
\begin{equation}  W_{\Sigma ls}(k_f)^{(2\pi\Sigma)}  = -W_{\Sigma ls}(k_f)^{
(2\pi\Sigma)}_{\rm Pauli} = {F^2 g_A^2 M_B m_\pi^2 \over 12 \pi^3 f_\pi^4} 
\bigg\{2\ln(1+u)- {2u+u^2 \over 1+u} \bigg\} \,, \end{equation} 
\begin{eqnarray}  U_{\Sigma ls}(k_f)^{(2\pi\Sigma)}_{\rm Pauli} &=& {F^2 g_A^2 
M_B m_\pi^2 \over 12 \pi^4 f_\pi^4}\Bigg\{ 6\sqrt{u} \arctan(2\sqrt{u})-2u
-{2\sqrt{u} \over \sqrt{1+u}}\ln(\sqrt{u}+\sqrt{1+u}) \nonumber\\ &&
-{3\over 2} \ln(1+4u)+\int_0^u \!\!dx \,{1+2u-2x \over (1+u-x)^2}\ln{(\sqrt{u}-
\sqrt{x})(1+u+x +2 \sqrt{ux}) \over  (\sqrt{u}+\sqrt{x})(1+u+x -2 \sqrt{ux}) }
\Bigg\} \,, \nonumber\\  \end{eqnarray} 
where now almost all integrals could be solved for the Pauli blocking 
correction. 

Summing up all calculated two-loop terms written in Eqs.(4-10) we show in 
Fig.\,2 the resulting complex $\Sigma$-nuclear spin-orbit coupling strength  
$U_{\Sigma ls}(k_f)+i\,W_{\Sigma ls}(k_f)$ as a function of the nucleon density 
in the region $0\leq \rho\leq 0.2\,$fm$^{-3}$ (corresponding to Fermi momenta
$k_f \leq 283\,$MeV). It is expected that higher-loop contributions related to 
pion-absorption on two nucleons, in-medium nucleon and pion self-energy
corrections etc. are small in this low-density region.   
The upper curve for the imaginary part $W_{\Sigma ls}(k_f)$
clearly displays the onset of the Pauli blocking effect at the threshold
density  $\rho_{\rm th}= 0.072\,{\rm fm}^{-3}$. It may come as a surprise that
Pauli blocking increases the magnitude of the negative imaginary part. But
going back to the original expression Eq.(3) one sees that the squared energy
denominator introduces as a weight function for imaginary part the derivative
of a delta-function. Therefore the usual argument of phase space reduction by
Pauli blocking becomes insufficient even for a qualitative estimate. At
normal nuclear matter density $\rho_0 = 0.16\,$fm$^{-3}$ (corresponding to a
Fermi momentum of $k_{f0} = 263\,$MeV) one finds for the total imaginary part
$W_{\Sigma ls}(k_{f0})=(-6.83-4.89)\,$MeVfm$^2 =-11.7$\,MeVfm$^2$, where the
second entry stems from Pauli blocking. The physics behind this imaginary
spin-orbit coupling strength is, of course, the $\Sigma N \to \Lambda N$ 
conversion process induced by $1\pi$-exchange. One can also see from Fig.\,2 
that the cusp effect in the imaginary part $W_{\Sigma ls}(k_f)$ causes some 
non-smooth behavior of the real part $U_{\Sigma ls}(k_f)$. The almost linear 
decrease with density gets interrupted at the threshold density  $\rho_{\rm 
th}= 0.072\,{\rm fm}^{-3}$. At saturation density one finds a ``wrong-sign'' 
$\Sigma$-nuclear spin-orbit coupling strength of $U_{\Sigma ls}(k_{f0}) = [(
-1.83-2.32)+(-18.21+2.43)]\,$MeVfm$^2=-19.9$\,MeVfm$^2$, where the individual 
entries correspond to respective terms written in Eqs.(4,6,8,10), in that 
order. It is somewhat larger than the ``wrong-sign'' spin-orbit coupling of a 
$\Lambda$ hyperon, $U_{\Lambda ls}(k_{f0}) =-15\,$MeVfm$^2$ \cite{lambdapot}. 
This is our major result: The second order $1\pi$-exchange tensor interaction 
generates sizeable ``wrong-sign'' spin-orbit couplings for the $\Lambda$ and  
the $\Sigma$ hyperon together. The negative sign in case of the $\Sigma$ 
hyperon is however less obvious, because the relevant loop integrals are
derivatives of six-dimensional principal value integrals (see Eq.(3)). As an
aside we note that in the chiral limit ($m_\pi=0)$ the $\Sigma$-nuclear
spin-orbit coupling strength changes to $U_{\Sigma ls}(k_{f0})+i\,W_{\Sigma 
ls}(k_{f0}) =(-25.0-13.0\, i)\,$MeVfm$^2$, with the real part coming now
entirely from the Pauli blocking corrections.  

\begin{figure}
\begin{center}
\includegraphics[scale=0.5]{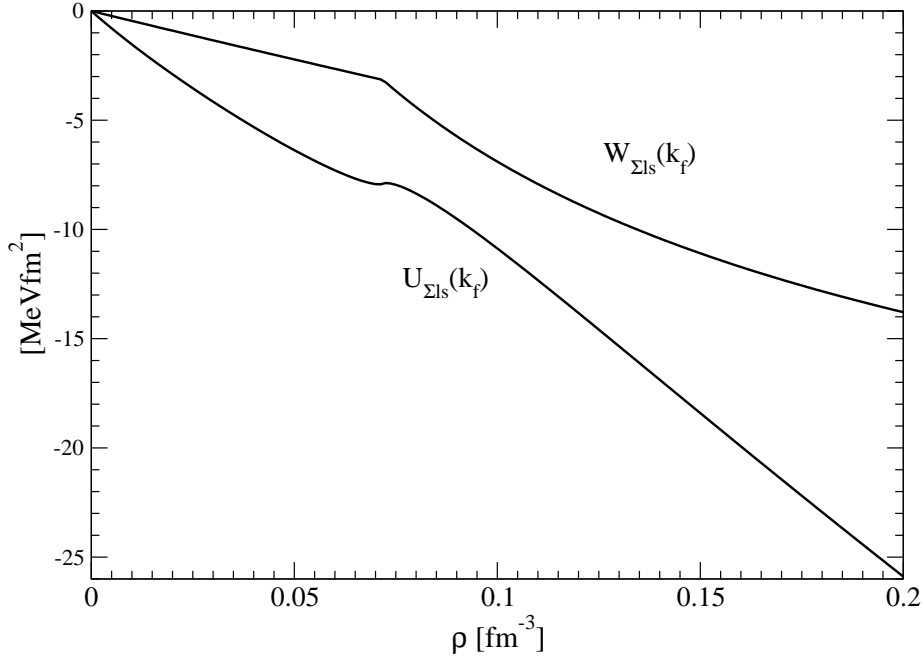}
\end{center}\vspace{-.2cm}
\caption{The complex-valued $\Sigma$-nuclear spin-orbit coupling strength 
$U_{\Sigma ls}(k_f)+i\, W_{\Sigma ls}(k_f)$ generated by iterated $1\pi
$-exchange as a function of the nucleon density $\rho= 2k_f^3/3\pi^2$. The 
imaginary part $W_{\Sigma ls}(k_f)$ originates from the conversion process
$\Sigma N \to \Lambda N$ induced by $1\pi$-exchange.} 
\vspace{-.3cm}
\end{figure}

It is expected that the additional $2\pi$-exchange effects of Ref.\cite{jorge}
including decuplet baryons in the intermediate state do not change the present 
results in a significant way. First, the additional mass splittings in the
energy denominators are so high that no new contribution to the imaginary part 
$W_{\Sigma ls}(k_f)$ is generated for $\rho\leq \rho_0$. Secondly, the 
approximate cancellation  between the contributions from $\Delta(1232)$ and
$\Sigma^*(1385)$ intermediate states works for $\Lambda$ and $\Sigma$ hyperons 
together, since it is based on different signs of spin-sums \cite{jorge}.

The short-range part of the $\Sigma$-nuclear spin-orbit interaction results
from a variety of processes, one of them being the $\omega$-exchange piece
presented in Eq.(2). Following Ref.\cite{lambdapot}, we relate the 
short-distance spin-orbit coupling of the $\Sigma$ hyperon to the one of the
nucleon as follows:   
\begin{equation}  U_{\Sigma ls}(k_f)^{(\rm sh)}=  C_{ls}{M_N^2 \over 
M_\Sigma^2} \,  U_{N ls}(k_f)^{(\rm sh)}\,. \end{equation}    
The factor $(M_N/M_\Sigma)^2 = 0.62$ results from the replacement of the
nucleon by a $\Sigma$ hyperon in these relativistic spin-orbit terms. The
coefficient $C_{ls}$ parameterizes the ratio of the relevant coupling 
constants. The expectation from the naive quark model would be $C_{ls}=2/3$. 
On the other hand, QCD sum rule calculations of $\Sigma$ hyperons in nuclear
matter \cite{jin} indicate that the Lorentz scalar and vector mean fields of a 
$\Sigma$ hyperon are similar to the corresponding ones of a nucleon,
i.e. $C_{ls} \simeq 1$. In case of the Lorentz scalar mean field, the QCD sum 
rule calculations are subject to uncertainties due to poorly known 
contributions from four-quark condensates. Ref.\cite{jin} concludes that due 
to a significant SU(3) symmetry breaking in nuclear matter the short-range 
spin-orbit term of a $\Sigma$ hyperon may be comparable to the one of a 
nucleon. For the further discussion we take for the short-range 
nucleonic spin-orbit coupling strength $U_{N ls}(k_f)^{(\rm sh)}= 3 \rho W_0/2 
= 30 \,{\rm MeVfm}^2 \rho/\rho_0$ with $W_0 =124\,{\rm MeVfm}^5$ the
spin-orbit parameter in the Skyrme phenomenology \cite{sly}. Employing $C_{ls} 
\simeq 1$, as indicated by the sum rule calculations, one  estimates the 
short-range $\Sigma$-nuclear spin-orbit coupling strength to $U_{\Sigma 
ls}(k_{f0})^{(\rm sh)} \simeq 18.6\,$MeVfm$^2$. This would lead to an almost
complete cancellation of the long-range component generated by iterated
one-pion exchange, resulting in a rather weak $\Sigma$-nuclear spin-orbit
coupling (admittedly with large uncertainties). Finally, we note that the 
long-range and short-range pieces are distinguished by markedly different 
dependences on the pion mass $m_\pi$ (or light quark mass $m_q \sim m_\pi^2$) 
and the density $\rho=2k_f^3/3\pi^2$. Therefore, there seems to be no double 
counting problem when adding long-range and short-range components.   

In summary, we have calculated in this work the $\Sigma$-nuclear spin-orbit 
coupling generated by iterated one-pion exchange with a $\Lambda$ or a 
$\Sigma$ hyperon in the intermediate state. We find from this unique 
long-range dynamics a sizeable ``wrong-sign'' spin-orbit coupling strength of 
$U_{\Sigma ls}(k_{f0}) \simeq -20$\,MeVfm$^2$.  When combined with estimates
of the short-range component a weak $\Sigma$-nuclear spin-orbit coupling
will result in total. Unfortunately, the prospects for an experimental check of
this feature are poor. The recently established repulsive nature of the
$\Sigma$-nucleus optical potential \cite{galneu} precludes a rich spectroscopy 
of heavy $\Sigma$-hypernuclei which could reveal spin-orbit splittings.        

Acknowledgments: I thank A. Gal and W. Weise for suggesting this work and for
informative discussions.

\end{document}